\documentstyle[12pt]{article}

\def\ai{\'{\i}}
\def\ite{\int_{t_1} ^{t_2}}

\def\itau{\int_{\tau_1} ^{\tau_2}}
\def\qb{\overline Q}
\def\pp{\pi_\phi}

\def\po{\pi_\Omega}
\def\om{\Omega}
\def\33{e^{3\Omega}}
\def\66{e^{6\Omega}}

\def\e-3{e^{-3\Omega}}

\def\ep{\epsilon}

\def\be{\begin{equation}}
\def\ee{\end{equation}}

\def\pb{\overline P}
\begin{document}

\baselineskip.33in

\centerline{\large{\bf Gauge invariance of parametrized systems}}

\centerline{\large{\bf and path integral quantization}}

\bigskip

\centerline{Hern\'an De Cicco$^{a,}$\footnote{Electronic address: decicco@cnea.gov.ar} and Claudio Simeone$^{b,c,}$\footnote{Electronic address: simeone@tandar.cnea.gov.ar}}

\bigskip

\noindent {\it a) Centro At\'omico Constituyentes,  Comisi\'on Nacional de Energ\ai a At\'omica, Av. del Libertador 8250 - 1429 Buenos Aires, Argentina.}

\noindent {\it b) Departamento de F\ai sica, Comisi\'on Nacional de Energ\ai a At\'omica,
 Av. del Libertador 8250 - 1429 Buenos Aires, Argentina.}

\noindent {\it c) Departamento de F\ai sica, Facultad de Ciencias Exactas y Naturales, Universidad de Buenos Aires,  Ciudad Universitaria, Pabell\'on I - 1428, Buenos Aires, Argentina.}

\vskip1cm

ABSTRACT

\bigskip

Gauge invariance of  systems whose  Hamilton-Jacobi equation is separable is improved by adding surface terms to the action fuctional. The general  form of these terms is given for some  complete solutions of the Hamilton-Jacobi equation. The procedure is applied to the relativistic particle  and toy universes, which  are quantized by imposing canonical gauge conditions in the path integral; in the case of empty models,  we first quantize the parametrized system called ``ideal clock'', and then we examine the possibility of obtaining the amplitude for the   minisuperspaces by matching them with the ideal clock. The relation existing between the geometrical properties of the constraint surface and the 
variables identifying the quantum states in the path integral is  discussed.
\vskip1cm


\newpage

{\bf 1. INTRODUCTION}

\bigskip

When the theory for a gauge system is given in the Hamiltonian formulation we obtain constraints $G_m$ which are linear and homogeneous in the momenta, plus a non vanishing Hamiltonian $H_0$ which is the total energy; for example, in the case of the electromagnetic field  the canonical momenta are the four quantities $F^{\mu 0}$; for $\mu=1,2,3$ we have the three components of the electric field, but for $\mu=0$ we have the constraint  $F^{0 0}=0$$^1$. The case of a parametrized system is different: the reparametrization invariance, associated to the fact that the evolution is given in terms of a parameter $\tau$ which does not have physical meaning, yields a Hamiltonian which vanishes on the physical trajectories of the system, that is, a constraint $G=H\approx 0$ which in most cases  is not linear and homogeneous in the momenta; for example,  minisuperspace models  have a constraint that is quadratic in the momenta (the reparametrization invariance of the models reflects the general covariance of the full theory of gravitation$^{2,3}$). 

When quantizing parametrized systems we are faced to the problem that these systems lack an important symmetry that ordinary gauge systems have:
under a gauge transformation defined by the parameters $\epsilon^m$ the action of a system with constraints $G_m$ changes by
\be\delta_\epsilon S=\left[ \epsilon^m (\tau )\left( p_i{\partial G_m\over\partial p_i}-G_m\right) \right]_{\tau_1} ^{\tau_2}. \ee
Then we have $\delta_\ep S=0$ for an ordinary gauge system, but, unless $\epsilon (\tau_1 )=\epsilon(\tau_2 )=0$, $\delta_\ep S\neq 0$ for a parametrized system, so that the last one does not have  gauge invariance at the end points. This has practical consequences, because to quantize the system one must impose gauge conditions, and these are restricted by the symmetries of the system: canonical gauges (those which are of the form $\chi(q,p,\tau)=0$)   would not be admissible for parametrized systems$^{4,5}$.

However, if the Hamilton-Jacobi (H-J) equation is separable, a parametrized system can be provided with gauge invariance over the whole trajectorie by improving its action functional with end point terms$^{6,7}$ which can be seen as the result of a canonical transformation which turns the system into an ordinary gauge system$^{7,8,9}$. In the present work  give the general form of the end point terms making the action invariant  and the appropriate gauge fixing procedure for several types of solutions of the H-J equation; then quantize some parametrized systems by means of the usual procedure for gauge systems. We obtain the Feynman propagator for the Klein-Gordon equation and the transition amplitude for toy universes. In the case of models with matter field we turn them into ordinary gauge systems and after a canonical gauge choice we obtain  simple expressions which show the separation between physical degrees of freedom and time. For empty models we proceed in two steps:  we first turn the parametrized system called ``ideal clock'' into an ordinary gauge system and quantize it with the usual path integral procedure of Fadeev and Popov; then the possibility of obtaining the transition amplitude for the minisuperspaces by matching them with the ideal clock is discussed. In particular, the restrictions arising from  the topology of the constraint surface are studied.

\vskip1cm

{\bf 2. PARAMETRIZED SYSTEMS}
\bigskip

Consider a mechanical system with canonical coordinates and momenta $(q^k, p_k)$. Its action functional reads
\be S[q^k,p_k] =\ite\left[ p_k {dq^k\over dt}-H_0(q^k, p_k)\right] dt,\ee
but as the dynamics remain unchanged if we add a total derivative of $t$ to the integrand, we can write
\be S[q^k,p_k] =\ite\left[ p_k {dq^k\over dt}-H_0(q^k, p_k)+R(t)\right] dt.\ee
We can give the evolution in terms of an arbitrary parameter $\tau$ by including the time $t$ among the canonical coordinates, so that the conjugate momentum $p_t$ appears. Now, if we want the action to lead to the same dynamics as  the original one does, the constraint $H=p_t+H_0 -R(t) \approx 0$  must be imposed; therefore,  the action for the  parametrized system with coordinates and momenta $(q^i, p_i)$ reads
\begin{eqnarray} S[q^i,p_i,N] & =& \itau\left[ p_t{dt\over d\tau}+p_k{dq^k\over d\tau}-N\left( p_t +H_0(q^k, p_k)-R(t)\right)\right] d\tau\nonumber\\                & = & \itau\left[ p_i{dq^i\over d\tau}-NH(q^i, p_i)\right] d\tau,\end{eqnarray}
where $N$ is a Lagrange multiplier. The usual canonical equations of motion for the   original coordinates and momenta $(q^k, p_k)$ should hold. Indeed, by varying the coordinates and momenta in (4) we obtain the canonical equations of motion
\begin{eqnarray}
{dq^i\over d\tau}&=& N{\partial\over\partial p_i}\left( p_t +H_0(q^k, p_k)\right)\nonumber\\
{dp_i\over d\tau}&=& -N{\partial\over\partial q^i}\left( H_0(q^k, p_k)-R(t)\right)\nonumber
\end{eqnarray}
which  give
\begin{eqnarray}
{dq^k\over dt}&=& {\partial\over\partial p_k}H_0(q^k, p_k)\nonumber\\
{dp_k\over dt}&=& -{\partial\over\partial q^k}H_0(q^k, p_k)\nonumber
\end{eqnarray}
 
If we eliminate the coordinates  and momenta $(q^k, p_k)$, we obtain a system with only one degree of freedom and one constraint, the ideal clock; its action functional is given by
\be
S[t, p_t,N]=\itau\left(p_t{dt\over d\tau}-NH\right) d\tau
\ee
with
\be H=p_t -R(t) \approx 0,
\ee
and the equations of motion for this system are
$${dt\over d\tau}=N,\ \ \ \ \ \ {dp_t\over d\tau}= N{\partial R(t)\over \partial t}.$$

The choice $R(t)={t^2\over 2}$ together with the transformation
\be
\tilde q^0=p_t,\ \ \ \ \ \  \tilde p_0=-t
\ee
yield the constraint
\be
\tilde H\equiv -{1\over 2}{\tilde p_0}^2+\tilde q^0\approx 0,
\ee
and it can be shown that there is a second transformation $(\tilde q^0,\tilde p_0)\ \to\ (\Omega,\po)$ leading to the constraint$^{10}$
\be
H\equiv -G(\om)\po^2+v(\om ) \approx 0, 
\ee
which can describe isotropic and homogeneous minisuperspaces with no matter field.

\vskip1cm

{\bf 3. GAUGE INVARIANCE AND END POINT TERMS}

\bigskip

When we quantize a constrained system we must choose a gauge which can be reached from any path in the phase space by means of gauge transformations leaving the action unchanged. Under the gauge transformation  
$$\delta_\epsilon q^i=\epsilon (\tau )[q^i,H], \ \ \  \delta_\epsilon p_i=\epsilon (\tau )[p_i,H], \ \ \  \delta_\epsilon N={d\epsilon\over d \tau}$$
the variation of the action (4) is
\begin{eqnarray}\delta_\epsilon S & = & \left[ \epsilon (\tau )\left( p_i{\partial H\over\partial p_i}-H\right) \right]_{\tau_1} ^{\tau_2} \nonumber\\
& = & \left[ \epsilon (\tau )\left( p_k{\partial H_0\over\partial p_k}-H_0+R(t)\right) \right]_{\tau_1} ^{\tau_2},\end{eqnarray}
which clearly does not vanish for general values of $\ep$ at the boundaries. To obtain a gauge invariant action we must add a surface term $B$ with the property
$$\delta_\epsilon B  =  -\left[ \epsilon (\tau )\left( p_i{\partial H\over\partial p_i}-H\right) \right]_{\tau_1} ^{\tau_2}.$$
The general form of the appropriate surface term is$^7$
\be
B=\left[ \qb^i\pb_i -W+Q^\mu P_\mu-f\right]_{\tau_1}^{\tau_2}
\ee
where: 1) $W(q^i,\pb_i)$ is a complete solution of the H-J equation
\be H\left( q^i,{\partial W\over\partial q^i}\right) =E \ee
with $\pb_0=E$ and $\pb_\mu=\alpha_\mu$ ($\alpha_\mu$ integration constants).

\noindent 2)  $(\qb^\mu,\pb_\mu)$ are (conserved) gauge invariant variables which result from a canonical trasformation generated by $W$.

\noindent 3) $f$ is the $\tau$-dependent generator function of a canonical transformation  $(\qb^\mu,\pb_\mu)\ \to\  (Q^\mu,P_\mu)$, where $Q^\mu$ and $P_\mu$ are non conserved gauge invariant variables. The transformation leads to a non vanishing Hamiltonian
$$K=NH+{\partial f\over\partial\tau}$$
with $H=P_0\approx 0$. 
The function $f$ must be chosen so that $B$ vanishes with a gauge choice defining $\tau=\tau(q^i)$ in order to ensure that the new gauge invariant action weighs the paths in the same way that the original action does.

The gauge invariant action is
\be
{\cal S}[q^i,p_i,N]=\itau \left[ p_i{dq^i\over d\tau}-NH\right] d\tau+B
\ee
and in terms of the new coordinates and momenta
\be
{\cal S}[Q^i,P_i,N]=\itau \left[ P_i{dQ^i\over d\tau}-NP_0-{\partial f\over\partial\tau}\right] d\tau,\ee
so that the action is  now that of a gauge system because it has a non vanishing Hamiltonian and the constraint $P_0\approx 0$ is linear and homogeneous in the momenta . According to the usual path integral procedure to quantize gauge systems the amplitude for the transition $|Q^i_1,\tau_1> \ \to\ |Q^i_2,\tau_2>$ is  given by
\be <Q^i_2,\tau_2 |Q^i_1,\tau_1 >=\int DQ^i DP_i DN \delta (\chi) |[\chi ,H]| e^{i{\cal S}}\ee
 where $|[\chi ,H]|$ is the Fadeev-Popov determinant, which makes the result independent of the gauge choice.

The new variables $(Q^i,P_i)=(Q^0,Q^\mu,P_0,P_\mu)$ fulfill
$$[Q^{\mu},P_0]=[Q^{\mu},H]=0$$
$$[P_{\mu},P_0]=[P_{\mu},H]=0$$
$$[Q^{0},P_0]=[Q^{0},H]=1.$$
The last equality is of great importance: when dealing with an ordinary gauge system with a constraint $G=P_0\approx 0,$ a global gauge condition $\chi=0$ is well defined if$^{11}$ 
\be
\det([\chi ,G])\not=0,\ee
so that $[Q^0,P_0]=1$ makes $Q^0$ the obvious choice for a gauge fixing function. From a different point of view, in the case of a parametrized system with a constraint $H=0$ we know that $T$ is a global phase time if$^{13}$
\be [T,H]>0;\ee
therefore $[Q^{0},H]=1$ means that $Q^0$ is a global phase time for the parametrized system. Any possible difficulty with the condition (16) is then the same as with the condition (17), and if we turn a parametrized system into  an ordinary gauge one and find a  global gauge condition to  quantize the gauge system, we also find a global phase time for the parametrized system as long as the transformation from $(q^i, p_i)$ to $(Q^i, P_i)$ is well behaved.

From (11) and (12) we see that to achieve our aim we need a complete solution of the $\tau$-independent  H-J equation. We shall give the surface terms together with the appropriate gauge fixing procedure for some forms of the complete solution $W$:

\smallskip

\noindent 1) If the system has two degrees of freedom and the solution $W$ is of the form
\be
W=A(q^0,q)A^{'} (\pb_0,\pb )+ A^{''} (q^0,q)\pb.\ee
we obtain
$$\qb^0={\partial                        W\over\partial\pb_0}=A(q^0,q){\partial
A^{'}(\pb_j)\over\partial\pb_0}.$$
Gauge fixation  must  define  an  hypersurface  in  the original configuration space. It holds if
\be
\chi\equiv    \qb^0    -    {\partial   A^{'} (\pb_j)\over\partial\pb_0}T(\tau
)=0,\ee
with $T(\tau)$ an arbitrary monotonic function, because then $$A(q^0,q)=T(\tau ).$$

End point terms vanish on the surface $\pb_0=0$, $\chi=0$ if we choose
\be
f(\qb ,P,\tau )=\qb P-T(\tau )A^{'} (\pb_0 =0,\pb =P).\ee
Then
$$Q\vert_{\pb_0 =0,\chi =0}=A''(q^0,q) $$
and the choice of $Q$ and $\tau$ is equivalent to the choice of $q^0$ and $q.$ This must be fulfilled  to be sure that $Q$ and $\tau$ define a point in the original configuration space.
 
The generator function of the two succesive transformations $x^i\rightarrow \overline X^i  
\rightarrow X^i $ can be writen
\be
Z=A(q^0,q)A^{'} (P_0,P)+A^{''} (q^0,q) P-T(\tau )A^{'} (P_0 =0,P),\ee
and on the constraint surface $P_0 =0$ the end point terms are
\begin{eqnarray}B & = & QP-Z\nonumber\\
 & = & {\partial Z\over\partial P}P-Z\nonumber\\
&  = & \left[    {\partial    A^{'}(P_0, P) \over\partial  P}P-A^{'}(P_0,P) \right]_{P_0  =0} 
\left( A(q^0,q)-T(\tau )\right).\end{eqnarray}
This form of the end point terms can be used for the parametrized (non relativistic) particle, the relativistic particle, and several systems that can be obtained from them.

\smallskip

\noindent 2) Another useful form of the generator $W$ for two degrees of freedom is 
\be
W=D(q^0,\pb_0,\pb )+C(q^i)\pb ,\ee
which yields
$$\qb^0 ={\partial W\over\partial \pb_0}={\partial D\over\partial \pb_0}.$$
The gauge condition
\be
\chi =\qb^0 -g(\pb ,T(\tau ))=0\ee
is equivalent, on the constraint surface, to  
$$q^0=T(\tau )$$
if the function
$$g(\pb ,T(\tau ))=\qb^0(q^0 =T(\tau ),\pb_0 =0,\pb )$$
is chosen.

The function $f$ making the end point terms vanish with this gauge choice is
\be
f(\qb ,P,\tau )=\qb P -D (T(\tau ),\pb_0 =0,\pb=P ),\ee
and then   
$$Q\vert_{\pb_0 =0,\chi =0}=C(q^i); $$
the choice of  $Q$ and $\tau$ is thus equivalent to that of  $q^0$ and $q.$
The two successive transformations can be seen as only one generated by
\be Z=W(q^i,\pb_i=P_i)- D (T(\tau ),\pb_0 =0,\pb=P )\ee
and the surface terms have the form
\begin{eqnarray}
B &=& {\partial Z\over\partial P}-Z\nonumber\\ 
& =& P\left[{\partial\over\partial P}\left( D (q^0, P_0 , P)-D(T(\tau),P_0,P)\right)\right]_{P_0=0}-\nonumber\\
 &  & \ \ \ \ \ \ \mbox{}-\left[ D (q^0, P_0 , P)-D(T(\tau), P_0,P)\right]_{P_0=0} .\end{eqnarray}
The action of several isotropic and homogeneous cosmological models with matter field can be improved  with gauge invariance at the boundaries by these  surface terms. 

\smallskip

\noindent 3) In a more general case, whenever $\left[ \qb^i \pb_i -W\right]_{\chi =0,
P_0 =0}$ depends on  only one of the momenta $P_\mu$, say $P_1$, the two successive canonical
transformations $x^i\to \overline X^i\to X^i$ can be obtained as  the  result  of  only  one transformation
generated by 
\be Z=W(q^i,P_i)-P_1 \int{\left[  \qb^i\pb_i - W\right]_{\chi =0,\pb_0 =0,\pb_1
=P_1}\over P_1 ^2} dP_1 .\ee
The endpoint terms associated to this generator are
\begin{eqnarray}B &=& Q^i P_i -Z \nonumber\\
& = & {\partial Z\over\partial P_i}P_i-Z\nonumber\\
& = & {\partial  W\over\partial  P_i}P_i  -\left[  \qb^i\pb_i -
W\right]_{\chi =0,\pb_0 =0,\pb_1 =P_1}  -W(q^i,P_i).\end{eqnarray}

\vskip1cm

{\bf 4. PATH INTEGRAL QUANTIZATION: EXAMPLES} 
\bigskip

{\bf A. Feynman propagator for the Klein-Gordon equation.}

\medskip

To illustrate our procedure let us begin by calculating the Feynman propagator for the Klein-Gordon equation
$$(-{\partial^2\over\partial t} + \nabla^2 +m^2)\psi=0.$$ 
The relativistic particle is a system analogous to gravitation in that it is invariant under local reparametrizations of time, and in the canonical formalism is described by a Hamiltonian constraint:
\be H={p_0}^2-p^2-m^2\approx 0\ee
(however, the relativistic particle cannot reproduce an important property of cosmological models, as it is the fact that the potential in the Hamiltonian constraint can change its sign). We shall obtain the propagator by computing the functional average of the Heaviside function $\theta(s),$ where $s$ is the proper time (to simplify the notation we  write only one spatial coordinate). 

The $\tau$-independent Hamilton-Jacobi (12) equation for this system, matching $E=\pb_0$ and $\alpha=\pb$, has the solution
$$W_\pm                                (x,x^0,\pb,\pb_0)                 =\pb
x\pm x^0{\sqrt{\pb^2+{\pb_0}+m^2}},$$
and corresponds to the case 1) of the preceding section. With our notation, 
$$A(q^0,q)=x^0,$$
$${A'}_\pm (\pb_0,\pb )=\pm {\sqrt{\pb^2+{\pb_0}+m^2}},$$
$$A^{''}(q^0,q)=x.$$ 
The generating functional for the transformation $(q^i,p_i)\to (Q^i,P_i)$ is given by 
\be Z=\pm x^0{\sqrt{\pb^2+{\pb_0}+m^2}}+xP\mp T(\tau){\sqrt{P^2 +m^2}},\ee
and on the constraint surface  the end point terms are
\be B\equiv \mp{m^2(x^0-T(\tau            ))\over\sqrt{P^2+m^2}},
     \ee
and vanish in the canonical  gauge $\tilde\chi\equiv x^0-T(\tau )=0$ which gives $\tau=\tau(q^i)$.
The new variables are given by
\begin{eqnarray*} Q^0 &=&\pm{mx^0\over{\sqrt{P^2+{P_0}+m^2}}}\\
Q &=& x\pm
{Px^0\over{\sqrt{P^2+{P_0}+m^2}}}\mp{PT(\tau )\over\sqrt{P^2+m^2}}\\
p_0 &=&\pm {\sqrt{P^2+{P_0}+m^2}}\\
p &= & P.\end{eqnarray*}
Therefore, the gauge invariant action reads
\be {\cal S} =\itau\left(  p_0{dx^0\over  d\tau}  +p{dx\over  d\tau}-NH\right)
d\tau \mp \  m^2\left[  {x^0-  T(\tau  )\over  \sqrt  {p^2+m^2}}\right]_{\tau_1}
^{\tau_2}\ee
and the amplitude for the transition $x_1\to x_2$ is given by
\begin{eqnarray} <x_2,x_2^0\vert x_1,x_1^0>&=&\int Dx^0 Dp_0  DxDpDN\delta  (\chi  )\vert  [\chi
,H]\vert \times\nonumber\\
& &\ \ \ \times\exp\left( i\itau\left(  p_0{dx^0\over  d\tau}  +p{dx\over  d\tau}-NH\right)
d\tau \right)\times\nonumber\\
& & \ \ \ \ \ \ \ \ \ \ \ \ \times\exp\left(\mp i\  m^2\left[  {x^0-  T(\tau  )\over  \sqrt{p^2+m^2}}\right]_{\tau_1}
^{\tau_2}\right) .\end{eqnarray}
The path integral can  now be   computed  in any canonical gauge; for any function $T$ we have $\vert  [\chi
,H]\vert =2|p_0| $. The integration on the multiplier $N$ yields a $\delta-$function of the constraint which can be writen as 
$$\delta (p_0^2-p^2-m^2)={1\over 2|p_0|}\delta (p_0-\sqrt  {p^2+m^2})+{1\over 2|p_0|}\delta (p_0+\sqrt  {p^2+m^2}).$$
In a $\tau-$independent gauge  we have $\theta (s)=\theta (x_2^0-{p_0(\tau_2)\over p_0(\tau_1)}x_1^0)$ for $p_0>0$ and  $\theta (s)=\theta (x_1^0-{p_0(\tau_1)\over p_0(\tau_2)}x_2^0)$ for $p_0<0$; then, in gauge $\chi\equiv
x^0=0$  we obtain 
\newpage

\begin{eqnarray} <x_2,x_2^0\vert\theta (s)\vert x_1,x_1^0>&=&\int DxDp\   \theta (x_2^0-{p_0(\tau_2)\over p_0(\tau_1)}x_1^0)\times\nonumber\\
& &\ \ \ \ \ \times\exp\left(i\itau p{dx\over  d\tau}
d\tau - i\  m^2\left[  {-  T(\tau  )\over  p_0}\right]_{\tau_1}
^{\tau_2}\right)\nonumber\\
& & +\int DxDp\   \theta (x_1^0-{p_0(\tau_1)\over p_0(\tau_2)}x_2^0)\times\nonumber\\
& &\ \ \ \ \ \times\exp\left(i\itau p{dx\over  d\tau}
d\tau +i\  m^2\left[  {-  T(\tau  )\over  p_0}\right]_{\tau_1}
^{\tau_2}\right),\end{eqnarray}
where
\begin{eqnarray}\itau p{dx\over  d\tau}
d\tau\pm  \left[ {m^2  T(\tau  )\over  p_0}\right]_{\tau_1}
^{\tau_2}&=&\itau  \left[    p{d\over   d\tau}\left(  x\mp{pT(\tau)\over  \sqrt
{p^2+m^2}}\right)\pm \sqrt  {p^2+m^2}{dT\over d\tau}\right]
d\tau \nonumber\\
&=&\itau \left[ P{dQ\over  d\tau}\pm  \sqrt  {P^2+m^2}{dT\over  d\tau  }\right]
d\tau .\end{eqnarray}
By skeletonizing the paths we obtain $N-1$ $\delta-$functions of the form $\delta(P_m-P_{m-1}),$ and as $P=p$ and the end point values of $Q$ are given by the gauge choice which makes the endpoint terms vanish, so that $Q(\tau_1)=x_1$ and $Q(\tau_2)=x_2,$ we finally obtain
\begin{eqnarray} <x_2,x_2^0\vert\theta (s)\vert x_1,x_1^0>&=&\theta (x_2^0-x_1^0)\int dp\exp\left( i \  p(x_2-x_1)- i\ p_0(x^0_2-x^0_1)\right)\nonumber\\
& &  + \theta (x_1^0-x_2^0)\int dp\exp\left( i \  p(x_2-x_1)+ i\ p_0(x^0_2-x^0_1)\right) ,\end{eqnarray}  
which is the Feynman propagator for the Klein-Gordon equation.

\bigskip

{\bf B. The ideal clock}

\medskip

A way to get a better understanding of the quantization of certain minisuperspaces is to recall that their Hamiltonian constraints can be  obtained by performing a canonical transformation on a mechanical system  which has been parametrized by including the time $t$  among the canonical variables. In particular, empty Friedman-Robertson-Walker (FRW) minisuperspaces  are obtained from the ideal clock, that is, a system whose only degree of freedom is the time.  Hence, we shall turn the ideal clock into an ordinary gauge system and compute its quantum transition amplitude by means of a  path integral in which canonical gauges are admissible. 

Following  the procedure given in the preceding section for the solutions of type 2), for the constraint $H=p_t-R(t)\approx 0$  we obtain 
$$W=D(t,\pb_0)=\pb_0 t+\int R(t) dt$$
and the generator function  is 
$$Z=-\pb_0 t-\int R(t) dt+\int R(\tau)d\tau.$$
On the constraint surface the appropriate boundary terms have the form
\begin{eqnarray}B&=&-Z\nonumber\\
&=& -\left[\int R(t)
dt  -\int R(\tau )d\tau \right]_{\tau_1} ^{\tau_2}.\end{eqnarray}
New and old variables are related by  $\qb^0={\partial W\over\partial\pb_0}={\partial W\over\partial E}=t$ and $\pb_0=P_0=E$. The gauge invariant action takes the form
\be{\cal S} = \itau
\left[ p_t  {dt  \over d\tau} -NH \right] d\tau-\int_{t(\tau_1)} ^{t(\tau_2)}R(t) dt +\itau
 R(\tau ) d\tau .\ee
Therefore, after integrating on the multiplier $N$ the amplitude for the transition $t_1\ \to\ t_2$ is given by
\begin{eqnarray} 
<t_2\vert t_1>& = & \int Dt Dp_t   \delta (\chi )\delta (p_t-R(t)) |[\chi,H]| \times\nonumber\\
& &\ \ \ \ \ \mbox{}\times \exp\left({i \itau 
p_t  {dt  \over d\tau} d\tau-i\int_{t(\tau_1)} ^{t(\tau_2)}R(t) dt +i\itau
 R(\tau ) d\tau }\right) .\end{eqnarray}
To perform the integration on $t$ and $p_t$ any canonical gauge choice is admissible; a very simple one is $\chi\equiv t=0$, which yields
\be
<t_2\vert    t_1>= \exp\left({i\itau    R(\tau)    d\tau
}\right) ,\ee
and thus the probability for the transition is 
\be |<t_2|t_1>|^2=1.\ee

In terms of the new variables $(Q^i,P_i)=(Q^0,P_0)$ the gauge invariant action reads
\be {\cal S}= \itau\left[  P_0{dQ^0\over  d\tau}-NP_0 -{\partial
f\over\partial\tau}\right] d\tau.\ee
In gauge $\tilde\chi\equiv Q^0-\tau =t-\tau =0 $ we have $\delta(\chi )=\delta (Q^0-\tau )$ and $|[\chi,H]|=[Q^0,P_0]=1$; hence, after integrating on $N$ we have
\begin{eqnarray}<t_2\vert t_1>&=&\int DQ^0 DP_0\delta(P_0)\delta (Q^0- \tau) \exp\left( i\itau\left[  P_0{dQ^0\over  d\tau} -{\partial
f\over\partial\tau}\right] d\tau \right)\nonumber\\
&=& \exp\left( -i\itau {\partial
f\over\partial\tau} d\tau \right) \nonumber\\
&=&  \exp\left(i\itau    R(\tau)    d\tau
\right). \end{eqnarray}

The result, of course, reflects that the only degree of freedom of our system can be used to  parametrize its evolution. This is clearly seen in the path integral (44), where the integration over the dynamical variables dissapears because the gauge choice $Q^0-\tau=0$ selects only one path in phase space. We should stress, however, that the result is independent of the gauge choice, as the path integral is gauge invariant. The point is not necessarily to impose this gauge condition,   but the possibility to choose it. 

In subsection 3.C we shall see how the ideal clock provides a way to examine the restrictions to our procedure which arise from the geometry of the constraint surface of minisuperspace models. 

\bigskip

{\bf C. Empty minisuperspaces}

\medskip

 The  Hamiltonian constraint

$$ H\equiv -G(\om)\po^2+v(\om ) \approx 0$$
with $\om\sim \ln a(\tau)$ ($a(\tau)$ the scale factor in the FRW metric) and $\po$ its conjugate momentum, which corresponds to  an empty minisuperpace, can be obtained from the constraint  of the ideal clock with $R(t)={t^2\over 2}$ by means of a suitable canonical transformation.
If we define$^{10}$
\be V(\om )=sign(v)\left( {3\over 2}\int\sqrt{|v|\over G} d\om\right)^{2/3}\ee
the canonical transformation is given by
\be \po= -t{\partial V(\om)\over\partial\om},\ \ \ \ \ \ 
p_t= V(\om ).\ee
On the constraint surface $p_t-t^2=0$ we obtain
\be t^2= V (\om), \ee 
and then we could obtain the amplitude for $\om_1 \to \om_2$ for the minisuperspace by means of 
the amplitude for $t_1 \to t_2$ if a parametrization in terms of $t$ is equivalent to a parametrization given by a function of $\om$ only.
Let us begin by  examining the behaviour of the potential $v(\om)$. The most general form of the potential for an empty FRW minisuperspace is
\be v(\om) = -k e^\om + \Lambda\33 ,
\ee
where $k=-1,0,1$ is the curvature and $\Lambda$ is the cosmological constant (we shall assume $\Lambda \geq 0$). Let us consider  first the simple models with $k=0$ or $\Lambda=0$. For $k=0$ (flat universe, non zero cosmological constant) we have
$$v_1 (\om)=\Lambda e^{3\om} ,$$
and for $\Lambda=0$, $k=-1$ (null cosmological constant,   open  universe) we have the potential
$$v_2 (\om)= e^{\om}$$
(the case $k=1,\ \Lambda=0$ is not possible). In both cases, as well as for the open $(k=-1)$ model with non zero cosmological constant, given $v$ and then $V$ we can obtain  $\om=\om(V)$ uniquely. This means that in this simple models $\om$ can play the role of time, i.e. the parameter in terms of which the evolution of the system is given. For $v_1$ and $v_2,$ as $\om\sim\ln a(\tau),$ it is clear that our procedure identifies the  scale factor of the metric  with the time $t.$ Observe that this is possible because when  the potential has a definite sign the constraint surface splits into two disjoint sheets identified by the sign of the momentum $\po$, so that $\om$ is enough to parametrize the evolution. The analogy between these systems and the ideal clock is therefore complete, and their quantization results then trivial.

The case $k=1$, $\Lambda > 0$ (closed model with non zero cosmological constant),
$$ v(\om) = -e^\om + \Lambda\33 , $$
 requires more care. This potential is not a monotonic function of $\om$, but it changes its slope when
\be
\om=\ln \left({1\over \sqrt{3\Lambda}}\right)\ee
where it has a minimun, so that for a given value of $v(\om)$ we  have two possible values of $\om.$  However, physical states lie on the constraint surface 
$$ -G(\om)\po^2 -e^\om + \Lambda\33 = 0,$$   
which restricts the motion to
$$\po=\pm\sqrt{{e^\om(\Lambda e^{2\om}-1)\over G}}.$$
As $G$ is a positive definite function of $\om$, the natural size of the configuration  space is given by
\be
\om\geq \ln\left({1\over \sqrt\Lambda}\right).\ee
Hence,   the potential does not change its slope  on the constraint surface, and it is always possible to obtain $\om=\om(v)$ in the physical region of the phase space. There is, however, a topological difference between this case and the formers: now, as the potential has not a definite sign, the constraint surface is no more topologically equivalent to two disjoint planes, making impossible to parametrize the system in terms of the coordinate $\om$ only; indeed, at the point $\om=\ln\left({1\over \sqrt\Lambda}\right)$ we have $v=0$ and $\po=0$, but $\dot \po\neq 0,$ so that the system can go from $(\om,\po)$ to $(\om,-\po)$. Then a gauge condition given in terms of $t$ does not give $\tau=\tau(\om)$ and we are not able to identify the path integral in the new variables with the amplitude $<\om_2|\om_1>$. In fact, we have that because the potential can vanish, $t$ must be defined as a function of $\om$ and $\po$ (see ref. 10, where  $t\sim -e^{2\om}\po$ is defined for this model; the authors discuss the identification of a global phase time, but, as we have said, it is closely related to the fixation of a globally good gauge).

\bigskip

{\bf D. Minisuperspaces with one true degree of freedom}

\medskip

The action for a FRW model  with matter reads$^2$

\be S=\itau\left( \pp\dot\phi+\po\dot\om-NH\right) d\tau\ee
where $\phi$ is the matter field, $\om\sim\ln a(\tau)$, $\pp$ and $\po$ are their conjugate momenta, and $N$ is a Lagrange multiplier enforcing the Hamiltonian constraint
\be H=G(\phi,\om)(\pp^2-\po^2)+v(\phi,\om ) \approx 0.\ee

We shall now give an example of our general procedure by quantizing a model with massless scalar field, zero cosmological constant but non zero curvature by means of a path integral which admits canonical gauge conditions.  We shall study the model described by the Hamiltonian constraint  
\be H={1\over 4}e^{-3\om}(\pp^2-\po^2)-ke^\om\approx 0,\ee
with  $k=-1$ (open universe). The constraint surface splits in
$$\po=\pm\sqrt{\pp^2+4 e^{4\Omega}},$$
and this restricts the analysis of the existence of a  global gauge to each one of this two disjoint surfaces. In the case $k=-1$ each surface is topologically equivalent to half a plane.
  
The $\tau$-independent H-J equation for this system is
\be\left({\partial W\over\partial \phi}\right)^2 - \left({\partial W\over\partial \om}\right)^2+4e^{4\om}=4Ee^{3\om},\ee 
and matching the integration constants $\alpha,\  E$ to the new momenta $\pb,\ \pb_0$ it  has the solution
$$W(q^i,\pb_i)=\pb\phi\pm\int        d\Omega\sqrt{\pb^2-4\pb_0
e^{3\Omega}+4 e^{4\Omega} },$$
which is of the form (23) if $q^0=\om$ and $q=\phi$. Following the procedure of section 3, the generator function for the transformation $(q^i,p_i)\ \to\ (Q^i,P_i)$ is given by
\begin{eqnarray}Z & = & P\phi\pm\int \sqrt{P^2-       4P_0e^{3\Omega}+4
e^{4\Omega}}d\Omega\mp\nonumber\\
&   &\  \ \ \ \ \ \ \ \     \  \mbox{} \mp\int^{T(\tau )} \sqrt{P^2+4    e^{4\Omega}}d\Omega.
\end{eqnarray}
The endpoint terms vanish in the canonical gauge
$$\chi\equiv\qb^0 -g(\pb, T(\tau ))=0,$$
which is equivalent to
\be \Omega = T(\tau ) \ee
 if we choose $ g(\pb, T(\tau ))=\qb^0(\om=T(\tau), \pb_0=0,\pb).$
On the constraint surface the physical degree of freedom $Q$ and the new non vanishing Hamiltonian $K$ are given by
$$Q=\left[ {\partial    Z\over\partial  P}\right] _{P_0=0}=\phi\pm{\partial\over\partial
P}\int_{T(\tau    )}    ^\Omega     \sqrt{P^2+4    e^{4\Omega}}d\Omega ,$$
\be K={\partial                Z\over\partial        \om}=\mp\sqrt{P^2+4
e^{4\om}}.\ee
As we have  $[\chi,H]=[Q^0,P_0]=1$  the gauge choice is globally good (there is no Gribov problem$^{11}$ because the gauge condition (56) corresponds to a plane $\om=constant$ at each $\tau$, and then ensures that if the surface $\chi=0$ at any $\tau$ intersected an orbit more than once, then at another $\tau$ we should have $[\chi,H]=0$) , and because the Poisson bracket is invariant under a canonical transformation   a global phase time exists for the system. As a global phase time $t$ must fulfill $[t,H]>0$ we must choose $\om=t$ if the  system evolves on the sheet given by $\po<0$, and $-\om=t$ if the system evolves on the sheet $\po>0$.
 
In the canonical gauge (56) the path integral for this minisuperspace takes the form
of that for a ``relativistic particle'' with  a $T-$dependent mass: after the integration on the multiplier $N$   and the variables $(Q^0,P_0),$ we have
\be<\phi_2,\Omega_2\vert \phi_1,\Omega_1>=\int  DQDP\exp\left(i\int_{T_1} ^{T_2}
\left[ PdQ\pm\sqrt{P^2+4 e^{4T}} dT\right] \right) ,\ee
where the boundaries are $T_1=\om_1$, $T_2=\om_2$, and the paths  in phase space go from $Q_1=\phi_1$ to $Q_2=\phi_2.$ The result shows the separation between true degrees of freedom and time yielding after a simple canonical gauge choice. 

\smallskip

Another completely analogous example is the flat $(k=0)$ FRW minisuperspace with cosmological constant $\Lambda >0$ described by the Hamiltonian constraint 
\be H={1\over 4}\e-3 (\pp^2-\po^2)+\Lambda\33 \approx 0.\ee
On the constraint surface the true Hamiltonian of the reduced system is given by  $K=\mp\sqrt{P^2+4\Lambda
e^{6\om}}$; after integrating on the variables $Q^0$ and $P_0$ and on the Lagrange multiplier, with the gauge choice (56) we obtain
\be <\phi_2,\om_2\vert\phi_1,\om_1 >=\int DQDP\, \exp\left( i\int_{T_1} ^{T_2} \left[ PdQ\pm\sqrt{P^2+4\Lambda e^{6T}}dT\right]\right).\ee
The end point values of $Q$ and $T$ are related with the original coordinates $\phi$ and $\om$ as before.

The fact that the resulting path integral is analogous to that for a relativistic particle (with variable ``mass'') makes simple to obtain the infinitesimal propagator (to get the finite propagator we should integrate on $Q$): the result for the case with potential $\Lambda e^{3\om}$ is$^{13}$
$$ <\phi_2,\om_1+\epsilon \vert \phi_1, \om_1>=\pm{\epsilon\Lambda^{1/2} e^{3\om_1}\over \sqrt{\epsilon^2-(\phi_2-\phi_1)^2}}H_1^{(1)}(2\Lambda e^{3\om_1}\sqrt{\epsilon^2-(\phi_2-\phi_1)^2}),$$
with $H_1^{(1)}$  the Hankel function defined in terms of the Bessel functions $J_1$ and $N_1$ as $H_1^{(1)}=J_1+iN_1$. This propagator fulfills the boundary condition $<\phi_2,\om_1+\epsilon \vert \phi_1, \om_1>\ \to\ \delta(\phi_2-\phi_1)$ when $\epsilon \to 0.$

Observe that we have succeeded in obtaining the amplitude for the transition between two states identified only by the  coordinates   because the potential is such that the constraint surface splits into two separate sheets, and then allows to parametrize the system in terms of the coordinate $\om$; this is not the general case (see below).

\vskip1cm

{\bf 5. DISCUSSION} 

\bigskip
The presence of a  constraint which is not homogeneous and linear in the momenta in the action of parametrized systems would make impossible, in principle, to quantize them by means of the path integral procedure for ordinary gauge systems. However, when the Hamilton-Jacobi equation for the system can be solved, we can provide the action with gauge invariance at the boundaries  by adding  end point terms, and then making canonical gauges admissible in the path integral. A general solution of the H-J equation is, in general, difficult to obtain. We have given the general form of the boundary terms and the appropriate gauge fixing procedure for several types of the solution.

The gauge choice is not only a way to avoid divergences in the path integral for a constrained system, but also a reduction procedure to physical degrees of freedom. Our procedure allows us to quantize simple cosmological models in such a way that by means of a simple canonical gauge choice the separation between true degrees of freedom and time is clear. The  resulting reduced system is governed by a time-dependent Hamiltonian,  reflecting  its evolution with changing ``external conditions'' which are the metric that has been matched with time.
 
 When we choose a gauge to perform the path integration, at each $\tau$ we select one point from each class of equivalent points; if we do this with a system which is pure gauge, i.e. that has only one  degree of freedom and one constraint, we select only one point of the phase space at each  $\tau .$ For example, the gauge choice $t-\tau=0$
tells us that the paths in the phase space can only go from $t_1=\tau_1$ at $\tau_1$ to $t_2=\tau_2$ at $\tau_2$; there is no other possibility. Hence, the probability that the system evolves from $t_1$ at $\tau_1$ to $t_2$ at $\tau_2$ cannot be anything else but unity. If the potential of an empty cosmological model has a definite sign and is such that a relation $V(\om)\leftrightarrow \om$ exists (see(47)), a gauge condition $t-g(\tau)=0$ leaves  only one possible value of $\om$ at each $\tau$ and the quantization of the minisuperspace is then trivial. 

The interest of the simple examples considered in the present work is  more conceptual than practical. Nevertheless,  our procedure  clearly shows the relation which exists between the geometry of the constraint surface and the possibility of identifying the quantum states in the path integral by means of only the original coordinates: we can ensure that the amplitude $<Q_2^\mu,\tau_2|Q_1^\mu,\tau_1>$ ($Q^0$ is a spurious degree of freedom for the gauge system)
is equivalent to the amplitude $<q_2^i|q_1^i>$ for  a minisuperspace  if the paths are weighted in the same way by ${\cal S}$ and $S$ and if $Q^\mu$ and $\tau$ define a point in the original configuration space, that is, if  a state $|Q^\mu,\tau>$ is equivalent to $|q^i>$; this happens if there exists a gauge such that $\tau=\tau(q^i)$, and such that the boundary terms $B$ vanish. But as we can see even in the simple case of the  empty FRW model with $k=1$ and $\Lambda >0,$ a gauge condition yielding  $\tau=\tau(q^i)$ cannot be defined  if the constraint surface does not split into two disjoint sheets, that is, if the potential does not have a definite sign. Hence, if we want to quantize the system by imposing canonical gauges in the path integral, in the most general case of a potential with a non definite sign we should admit the possibility of identifying the quantum states in the original phase space not by $q^i$ but  by a complete set of functions of the coordinates and momenta $q^i$ and $p_i$.

\vskip1cm

REFERENCES

\bigskip
\noindent
1.  P.  A.  M.   Dirac,  {\it Lectures  on  Quantum  Mechanics,}  Belfer
Graduate School of Science, Yeshiva University, New York (1964).

\noindent 2.  J.    J.   Halliwell, in {\it Introductory Lectures  on  Quantum
Cosmology,}  Proceedings  of  the Jerusalem Winter School on Quantum Cosmology
and Baby Universes. edited by  T. Piran, World Scientific, Singapore (1990).

\noindent 3. A. O. Barvinsky, Phys. Rep. {\bf 230}, 237 (1993).

\noindent 4. C. Teitelboim, Phys. Rev. D {\bf 25}, 3159 (1982).

\noindent 5.  J. J. Halliwell, Phys. Rev. D {\bf 38}, 2468 (1988).

\noindent 6. M.   Henneaux, C.  Teitelboim and J.  D.  Vergara, Nucl. Phys. B 
{\bf 387}, 391 (1992).

\noindent 7. R. Ferraro and  C. Simeone,  J. Math. Phys. {\bf 38}, (1997). 

\noindent 8. C. Simeone, J. Math. Phys. {\bf 39}, 3131 (1998). 

\noindent 9. H. De Cicco and C. Simeone, submitted to Gen. Rel. Grav.
     
\noindent 10.  S.  C.   Beluardi and R.  Ferraro, Phys. Rev. D {\bf 52}, 1963 (1995).

\noindent 11. M.  Henneaux and C.  Teitelboim, {\it Quantization of Gauge Systems,}
Princeton University Press, New Jersey (1992).

\noindent 12. P. H\'aj\ai cek, Phys. Rev. D {\bf 34}, 1040 (1986).

\noindent 13. R. Ferraro, Phys. Rev. D {\bf 45}, 1198 (1992).

\end{document}